# Totally Homogeneous Networks

Dinghua Shi, Shanghai University, Shanghai 200444, China

Linyuan Lü, University of Electronic Science and Technology of China, Chengdu 611731, China

Guanrong Chen, City University of Hong Kong, Hong Kong SAR, China

**Abstract**

In network science, the non-homogeneity of node degrees has been a concerned issue for study. Yet, with the modern web technologies today, the traditional social communication topologies have evolved from node-central structures to online cycle-based communities, urgently requiring new network theories and tools. Switching the focus from node degrees to network cycles, it could reveal many interesting properties from the perspective of totally homogeneous networks, or sub-networks in a complex network, especially basic simplexes (cliques) such as links and triangles. Clearly, comparing to node degrees it is much more challenging to deal with network cycles. For studying the latter, a new clique vector space framework is introduced in this paper, where the vector space with a basis consisting of links has the dimension equal to the number of links, that with a basis consisting of triangles has the dimension equal to the number of triangles, and so on. These two vector spaces are related through a boundary operator, e.g., mapping the boundary of a triangle in one space to the sun of three links in the other space. Under the new framework, some important concepts and methodologies from algebraic topology, such as characteristic number, homology group and Betti number, will have a play in network science leading to foreseeable new research directions. As immediate applications, the paper illustrates some important characteristics affecting the collective behaviors of complex networks, some new cycle-dependent importance indexes of nodes, and implications for network synchronization and brain network analysis.

## Introduction

Network science has gained popularity due to its great achievements in the past twenty years, where small-world networks [1] are built from nearest-neighbor regular networks through rewiring, presenting two significant characteristics of short average path-length and large clustering coefficient, while scale-free networks [2] are modeled based on random networks [3], possessing a scale-free power-law node-degree distribution. The three fundamental concepts in network science—average path-length, degree distribution and clustering coefficient—correspond to three basic structures: chain, star and cycle. Clustering coefficient is calculated based on triangles, but cycle depends on many more, so there should be a large class of complex networks that have prominent cyclic structures that go beyond the typical small-world and scale-free models. Indeed, there is one, recently discovered [4], referred to as *totally homogeneous networks*, which was called '$(k,g,l)$-homogeneous network' [4], where $k$ is the node degree variable, $g$ is the girth variable (length of the smallest cycle of a node) and $l$ is the path-sum variable (sum of all path-lengths of a node to other nodes). This new class of networks were coined as the type of networks that have optimal synchronizability if it has a maximum $g$ and a minimum $l$, which is a particularly important type of regular networks that optimize the network synchronizability [4], as well as controllability and robustness against attacks. Despite their usefulness in various applications (as further discussed below), it is technically challenging to find such networks，since they require all nodes to have the same degree, the same girth and the same path-sum. Yet small-sized totally homogeneous sub-networks ubiquitously exist in all complex networks, typically triangles and smallest $k$-cavities (to be further discussed below). Notably, totally homogeneous networks have natural symmetries, signifying their importance in network theory and applications, therefore should be more carefully investigated.

A network usually has fully-connected sub-networks (i.e., complete sub-graphs) like triangles and tetrahedrons, called simplexes in topology or cliques in graph theory, which are special totally homogeneous (sub)networks. Here, a node is a 0-clique, a link is a 1-clique, a triangle is a 2-clique, a tetrahedron is a 3-clique, and so on. Other than cliques and nearest-neighbor regular (sub)networks, it is important to find those non-fully-connected and yet linearly-independent cavities as well as other sync-optimal totally homogeneous (sub)networks, which are known to be very important in brain functional networks and big-data analysis as will be further discussed below. Typical examples of totally homogeneous (sub)networks, namely, a fully-connected network, a smallest $k$-cavity network, a nearest-neighbor regular network and a sync-optimal network, are shown in Fig. 1.

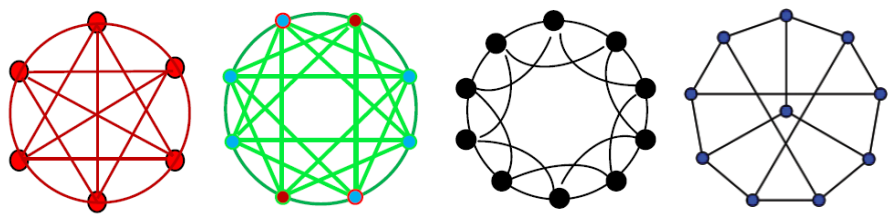

Figure 1. A 6-node fully-connected network; an 8-node smallest 3-cavity network; a 10-node nearest-neighbor regular network; a 10-node sync-optimal network

A network ~~is~~ can be represented by ~~the~~ an adjacency matrix, in which element 1 means two corresponding nodes are connected by a link, while 0 means not. Therefore, node degree is of the most importance for a network and, as a result, the emergence of giant nodes (hubs) signifies the key role of stars in scale-free networks, which has been a main research subject in the past.

By examining cycles, it is easy to find that cliques, e.g. links and triangles, constitute the backbone of a network. Therefore, a new clique vector space framework in the form of a sequence of vector spaces defined on the binary field is introduced. Let $C_0$ be the vector space with a basis consisting of all nodes, $C_1$ be one with a basis of all links (edges), and $C_2$ be the one with a basis of all triangles, and so forth. Noting that the three links of a triangle in $C_2$ are links in $C_1$, a boundary operator is defined between these two vector spaces, so that the two vector spaces connected by the boundary operator can be described and analyzed using a boundary matrix. It will be seen that boundary matrices have richer mathematical contents and are useful tools for advanced studies. For instance, one can introduce the concept of linear dependence for chain vectors and for cycle vectors, as well as equivalence between cycles, so that the resulting chain groups and cycle groups (including boundary groups and homology groups) can be investigated by using advanced mathematical tools from algebraic topology and group theory. It can be foreseen that such a clique vector space framework will enrich the studies of network science and meanwhile bring up more opportunities to the field.

Using tools from algebraic topology can help find invariants in networks. The most well-known one is perhaps the Euler characteristic number, which equals the alternative sum of all simplexes in the network, namely, the number of nodes minus the number of links then plus the number of triangles and then minus the number of tetrahedrons … until no more to add or subtract [5]. Another important invariant is the Betti number, which is the total number of linearly independent cavities [5], defined as follows: the order-0 Betti number is the number of connected sub-networks, order-1 Betti number is the number of linearly independent 1-cavities (non-triangular cycles), and so on, as will be further illustrated by examples below. The Euler-Poincaré formula connects these two indexes together [5]: the alternative sum of simplexes equals the alternative sum of the Betti numbers. Besides, the number of links in a spanning tree of a connected network equals the number of all nodes minus 1, which is the rank of the chain group in the network [6]. Moreover, the rank of the cycle group, namely the number of linearly independent triangular and non-triangular cycles, equals the number of links minus the number of nodes and then plus 1 [6].

Cycles have been a main research subject in graph theory [6], while algebraic topology is an important branch in mathematics [5]. In the current literature of network science, however, investigations involving both network cycles and algebraic topology are quite rare. Nevertheless, there are some significant progresses: the finding of a criterion for network synchronizability (2002)[7]; the introduction of (sync-optimal) totally homogeneous networks (2013) [4]; the study of signals and noise in brain activities (2017) [8]; the discovery of cliques and cavities in brain functional networks (2018) [9]; the search for basic cycles in measuring the importance of a node and cycle index in spreading over WeChat, and their relationships with hyper-networks (2019) [10].

This paper presents the mathematical description of a new framework of clique vector spaces, firstly introducing related concepts of various cycles, secondly defining a sequence of clique vector spaces associated with boundary operators, and finally discussing chain group, cycle group, boundary group and homology group. Then, it shows how to utilize boundary matrices to calculate linearly independent cycles and cavities of different orders, as well as their characteristic indexes, in a complex network. Furthermore, by examining the key factors that affect the collective behaviors of a network, it demonstrates the important role of totally homogeneous sub-networks in a complex network. Finally, it briefly discusses some applications of algebraic topology in network science, indicating a couple of future research topics.

**Description of the New Mathematical Framework**

Investigating network cycles is much more difficult than examining node degrees, since cycles have many variants such as higher-order cycles, linearly (in)dependent cycles, and redundant cycles that contain smaller circles. A simple cycle is intuitively clear, which is a closed path starting from a node and returning to the same node after traversing some other nodes. The smallest cycle is the triangle, which is also a clique, called 2-clique, and the vertexes and edges of a tetrahedron together constitute a clique, called the 3-clique. The 3-clique is not a cycle in the usual sense, which is called a send-order cycle, or simply a 2-cycle. Similarly, 4-clique is a 3-cycle, 5-clique is a 4-cycle, and so on. Yet, 0-clique and 1-clique are not cycles. The concept of linearly (in)dependent cycles or cavities will be introduced through the description of a the new framework next.

The new framework of clique vector spaces in the binary field and their associated boundary operators are defined as follows. Let $C_k$ be the vector space with a basis consisting of $k$-cliques, with dimension $m_k$ equal to the number of the $k$-cliques. The Euler characteristic number $\chi = m_0 - m_1 + m_2 - \cdots$ . All the vectors in $C_k$ are subsets of some $k$-cliques, with the empty set being the zero vector, denoted as $\emptyset$ or 0. In the binary field, there are only two elements, 0 and 1, with $1 + 1 = 0$. The addition between two vectors $c$ and $d$ is defined by set operations as $c + d = (c \cup d) - (c \cap d)$.

Define a boundary operator $\partial_k: C_k \to C_{k-1}$ to connect two successive vector spaces in the following way: denote triangle $(1,2,3) \in C_2$ by $\sigma_{123} \in C_2$, which has boundaries $(1,2), (2,3), (3,1)$, and define $\partial_2(\sigma_{123}) = \sigma_{12} + \sigma_{23} + \sigma_{31}$, where the "+" operation is performed in the binary field. For the two end nodes of the boundary (1,2), node 1 and node 2, one has $\partial_1(\sigma_{12}) = \sigma_1 + \sigma_2$. Again, by using additions in the binary field, one obtains

$$\partial_1(\partial_2(\sigma_{123})) = \partial_1(\sigma_{12} + \sigma_{23} + \sigma_{31}) = \sigma_1 + \sigma_2 + \sigma_2 + \sigma_3 + \sigma_3 + \sigma_1 = 0.$$

Since $C_1$ is a vector space consisting of links, its elements are called 1-chains. If, for $l \in C_1$, the chain satisfies $\partial_1(l) = 0$, then it is called a 1-cycle. Thus, a $k$-cycle $l$ is defined by $\partial_k(l) = 0$. Further, define the kernel of $C_k$ by $\ker(\partial_k) = \{l \in C_k \mid \partial_k(l) = 0\}$, denoted as $Z_k$, which is called the null space or kernel space. Moreover, $\text{im}(\partial_{k+1}) = \{\partial_{k+1}(l) \mid l \in C_{k+1}\}$ is the image of $C_{k+1}$ mapping to $C_k$, denoted as $Y_k$, called the image space on $C_k$. Clearly, both $Z_k$ and $Y_k$ are subspaces of $C_k$. Since $\partial_k(\partial_{k+1}) = 0$, one has $\text{im}(\partial_{k+1}) \subseteq \ker(\partial_k)$. The containment relationships of these subspaces are shown in Fig. 2.

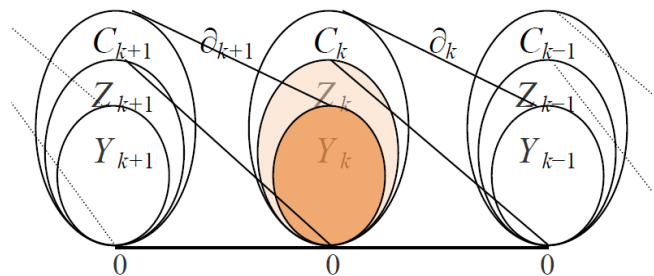

Figure 2. Relationships of a sequence of vector spaces in a network, with boundary operators and associated subspaces.

To study the linear dependence, boundary matrices are introduced. For instance, in $C_1$, one can define a node-link matrix $B_1$, in which an element is 1 if the node is in the link correspondingly; otherwise, it is 0. Similarly, define a link-face matrix $B_2$ on $C_2$, in which an element is 1 if the link is on the face correspondingly; otherwise, it is 0. Through elementary row (column) transformations in the binary field, one can calculate the rank $r_k$ of the boundary matrix $B_k$, which is the number of linearly independent vectors in ~~space~~ $C_k$. Moreover, all $k$-chains in $C_k$ constitute an Abel group, called a chain group, with the empty set being the zero element. The number of generating element of a chain group is the rank $r_k$ of the boundary matrix $B_k$. Likewise, all the $k$-cycles of $\ker(\partial_k)$ and all $k$-boundaries of $\text{im}(\partial_{k+1})$ each forms an Abel group, called cycle group $Z_k$ and boundary group $Y_k$, with ranks $m_k - r_k$ and $r_{k+1}$, respectively. Two $k$-cycles $c$ and $d$ are said to be equivalent, denoted as $c \sim d$, if $c + d$ is a boundary of a $(k + 1)$-chain. All equivalent cycles constitute an equivalent class. By definition, if $b \in Y_k$, then $b \sim \emptyset$. Decomposing the cycle group $Z_k$ via the boundary group $Y_k$ yields a homology group $Z_k/Y_k$, with rank equal to the Betti number $\beta_k = m_k - r_k - r_{k+1}$. Elements of this homology group are cycle-equivalent classes, called $k$-cavities.

Next, the above-introduced concepts and algorithm are illustrated by a simple network.

See *Supplementary Information*, Section 1 for detailed computations.

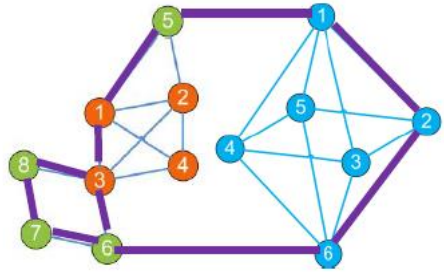

Figure 3. A network with 14 nodes and 26 links.

The network in Fig. 3 has 13 triangles and 1 tetrahedron, thus its Euler characteristic number is $\chi = 14 - 26 + 13 - 1 = 1 - 2 + 1 = 0$, where the Betti numbers are calculated in *Supplementary Information*, Section 1. Since $\beta_1 = 2$, the homology group $Z_1/Y_1$ has rank 2 and order 4, with 4 cycle-equivalent classes: empty set $\emptyset$ (including 13 triangles), two 1-cavities, and their sum. The key is to search for these cliques and cavities. For distinction below, the right-side node numbers are marked by apostrophes.

**Finding a spanning tree of the network**. Links (1,2) (1,3) (1,4) (1,5) (3,6) (3,8) (6,7), (5,1'), (1',2') (1',3') (1',4') (1',5') (2',6') constitute a spanning tree, which is not unique.

**Searching for linearly independent cycles of the network**. Joining other links to the spanning tree form linearly independent cycles, with total number equal to the number of links minus the number of nodes and then plus 1. In so doing, linearly dependent cliques and equivalent cavities with the same length in the network should also be included.

 Joining links (2,3) (2,4) (2,5) yields (1,2,3) (1,2,4) (1,2,5), respectively.
 Joining link (3,4) yields (1,3,4), and backtracking to (2,3,4) **(1,2,3,4)** (3-clique).
 Joining link (7,8) yields (3,6,7,8).
 Joining link (6,6') yields (1,5,1',2',6',6,3), and backtracking to (2,5,1',2',6',6,3).
 Joining links (2',3') (3',4') (4',5') (5',2') yields (1',2',3') (1',3',4') (1',4',5') (1',5',2').
 Joining link (3',6') yields (2',3',6'), and backtracking to (1,5,1',3',6',6,3) (2,5,1',3',6',6,3).
 Joining link (4',6') yields (3',4',6'), and backtracking to (1,5,1',4',6',6,3) (2,5,1',4',6',6,3).
 Joining link (5',6') yields (4',5',6'), and backtracking to (5',2',6') (1,5,1',5',6',6,3) (2,5,1',5',6',6,3) **(1',2',3',4',5',6')** (2-cavity).

Here, those cycles without underlines are linearly independent 1-cycles. Obviously, all 1-cycles and 2-cycles are totally homogeneous sub-networks.

To this end, it should be clear that the above analytic and computational methods can be extended to directed and weighted complex networks, even multi-layered networks.

Next, the invariants and linearly independent cycles of the four totally homogeneous networks shown in Fig. 1 are computed.

(1) The 6-node fully-connected network (5-clique)
   Characteristic number:   $\chi = \sum_{i=0}^{5}(-1)^i C_6^{i+1} = 1$
   Betti numbers:   $\beta_0 = 1, \beta_1 = \cdots = \beta_5 = 0$
   Linearly independent cycles:   $15 - 6 + 1 = 10$

(2) The 8-node smallest 3-cavity is a closed structure enclosed by tetrahedrons
   Characteristic number:   $\chi = 8 - 24 + 24 - 8 = 0$
   Betti numbers:   $\beta_0 = 1, \beta_1 = \beta_2 = 0, \beta_3 = 1$
   Linearly independent cycles:   $24 - 8 + 1 = 17$

(3) The 10-node nearest-neighbor regular network
   Characteristic number:   $\chi = 10 - 20 + 10 = 0$
   Betti numbers:   $\beta_0 = 1, \beta_1 = 1, \beta_2 = 0$
   Linearly independent cycles:   $20 - 10 + 1 = 11$

(4) The 10-node sync-optimal network
   Characteristic number:   $\chi = 10 - 15 + 0 = -5$
   Betti numbers:   $\beta_0 = 1, \beta_1 = 6, \beta_2 = 0$
   Linearly independent cycles:   $15 - 10 + 1 = 6$

The four representative totally homogeneous networks shown in Fig. 1 have characteristic numbers as follows:

   Characteristic number of a simplex:   $\chi = 1$
   Characteristic number of a smallest $k$-cavity:   $\chi = 1 + (-1)^k$
   Characteristic number of a regular network:   $\chi = 0$
   Characteristic number of a sync-optimal network of degree 3:
       $\chi = -n/2, \ n = 6, 8, 10, \ldots$

It is observed that the characteristic number of a smallest $k$-cavity oscillates between 0 and 2 depending on whether the $k$ is odd or even, perhaps due to the various dynamic properties of such networks ("topological cavities of different dimensions, around which information may flow in either diverging or converging patterns"[9]). It is also noted that the characteristic number of a sync-optimal network is always negative.

**Main Factors Affecting Collective Behaviors of a Complex Dynamical Network**

The seminal paper [1] on small-world networks studies the collective behaviors of a dynamical network. It points out that regular networks have relatively large clustering coefficients but their average path-lengths are generally quite long, and that random networks are opposite. So, it concludes that both are not good for collective dynamical behaviors such as information spreading and multi-agents synchronization. Thereby, it recommends a small-world network model that has both advantages of large clustering coefficients and short average path-lengths.

Now, focusing on cyclic structures in small-world networks reveals some interesting phenomena that have not been observed or emphasized before.

- **Network synchronization**—Characteristic number is key

In the study of optimal synchronizability of complex networks, it was found [4] that the totally homogeneous network with equal node degree, long girth and short path-sum, is the best. Here, the synchronizabilities of four typical networks shown in Fig. 4 are compared: regular network, small-world network, random network and sync-optimal totally homogeneous network, in which the small-world network is created through random rewiring [1] while the sync-optimal network through deterministic rewiring [4]. All these sample networks are connected with 20 nodes and 40 links without tetrahedrons.

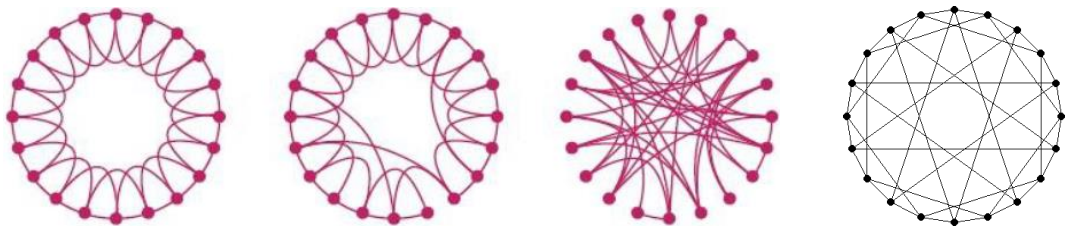

Figure 4. Nearest-neighbor regular network, small-world network,
random network and sync-optimal network

Data about these four sample networks are summarized below.

(1) Nearest-neighbor regular network
Triangles: 20
Characteristic number:  $\chi = 20 - 40 + 20 = 0$

Betti number:   $\beta_0 = 1$, $\beta_1 = 1$, $\beta_2 = 0$

(2) Small-world network

Triangles:   17

Characteristic number:   $\chi = 20 - 40 + 17 = -3$

Betti number:   $\beta_0 = 1$, $\beta_1 = 4$, $\beta_2 = 0$

(3) Random network

Triangles:   15

Characteristic number:   $\chi = 20 - 40 + 15 = -5$

Betti number:   $\beta_0 = 1$, $\beta_1 = 6$, $\beta_2 = 0$

(4) Sync-optimal network

Triangles:   0

Characteristic number:   $\chi = 20 - 40 + 0 = -20$

Betti number:   $\beta_0 = 1$, $\beta_1 = 21$, $\beta_2 = 0$

To compare their synchronizabilities, the eigenvalues of their Laplacian matrices are calculated [4,7]: the nearest-neighbor regular network has spectral gap (smallest nonzero eigenvalue) 0.4799 with eigen-ratio (smallest nonzero eigenvalue versus largest eigenvalue) 0.0769; the small-world network has spectral gap 0.5035 and eigen-ratio 0.0714; the random network has spectral gap 0.7947 and eigen-ratio 0.0812; the sync-optimal network has spectral gap 2.0000 and eigen-ratio 0.2982.

These results depict that the network synchronizabilities of the regular network, small-world network, random network and sync-optimal network are increasing successively. In particular, it shows that the key factor affecting the network synchronizability should be the Euler characteristic number: the smaller the characteristic number, the better the synchronizability for networks of the same size. Furthermore, the characteristic number depends on both 2-cliques and 1-cavities: having less 2-cliques but more 1-cavities, the characteristic number will be smaller, hence the network is easier to synchronize. These data are consistent with the previous observations. Note also that the clustering coefficient depends on the number of triangles, so a larger clustering coefficient means more 2-cliques are involved, and consequently the network synchronizability will become worse. Relatively to a larger clustering coefficient, a shorter average path-length is more important for better network synchronization.

- **Network spreading**—Totally homogenous networks are better

Now, consider information or disease spreading on the four networks shown in Fig. 4, all based on the cyclic SIR model [10] . This cycle-based SIR model differs from the conventional SIR model [11] in that the nodes belonging to the same cycle can always transmit information or disease even if they are not directly connected to each other.

In the performed simulations, successively selecting every node as the source and,

with spreading probability 0.06, processing the information though nodes with S (susceptive), I (infected) and R (recover) states, until no more infected nodes remaining in the network. Then, after 100 runs, the average number of recovered nodes was recorded. The final results were Network 1: 2.656 nodes, Network 2: 2.441 nodes, Network 3: 1.349 nodes, Network 4: 4.111 nodes (see *Supplementary Information*, Section 2 for more details).

These simulation results clearly show that the random network is the worst, the small-world network is not as good as the regular network, and the totally homogeneous network is the best.

**Other Promising Applications Based on Cycles**

- **Cycle-based importance indexes of nodes**—cycle number and cycle ratio

There are many indexes for measuring the importance of a node in a network [12], but there does not seem to have any based on cycle-structure.

Two new concepts of cycle number and cycle ratio are recently introduced [10] for measuring the importance of nodes. A node may have several smallest cycles, called smallest basic cycles of the network. All non-redundant cycles are called basic cycles. The cycle number of a node is defined as the total number of basic cycles that pass this node. The cycle ratio of a node $i$ is then defined to be the sum of the proportions of this node $i$ appearing in the basic cycles of all those nodes that are contained in the basic cycles of this node $i$.

To evaluate the performances of the indexes' abilities of measuring the node importance in a network, three existing indexes (degree, H-index and coreness) and the two new indexes (cycle number and cycle ratio) are compared [10] to study their effects on the connectivity and spreading over a network. For connectivity, all nodes are ranked according to their importance measured by an index and some nodes are removed, then a portion of nodes in a giant surviving sub-network is computed, and finally their relationships are plotted for comparison. For nodes with a same index value, randomly rank them one after another. Simulation demonstrates that intentional attack according to the cycle ratio ranking is more effective. For spreading, choose initial node according to the importance ranking, from high to low, as the source node. The SIR spreading process is then performed, until no infected node remains. After 100 runs, the average number of recovered nodes is recorded and the Kendall's tau correlation coefficient [13] is calculated. In evaluating spreading performance, due to the cyclic structure, the spreading matrix is used instead of the adjacency matrix. Here, the spreading matrix is introduced from WeChat data, which means that two nodes in the same group can communication even if they do not know each other. Results show that infected nodes spread very well according to the cycle number ranking on the cycle-based SIR model.

- **Relation of network and hypernetwork**—Studying hypernetworks

In classical graph theory, one link can only connect two nodes. In reality, however, one link could be shared by multiple nodes. Such a link is called a hyperlink. A network consisting of nodes and hyperlinks is called a hypernetwork. There exists correlation of an ordinary network and a hypernetwork [10]. By viewing a basic cycle of a node as a hyperlink, an ordinary network can be converted to a hypernetwork, as shown in Fig. 5. The questions are whether the reverse can be performed and, if so, whether the reverse process preserves all information. Since ordinary networks are special cases of hypernetworks, it is clear that generally the answers are no. But, this does not exclude particular situations. In fact, for a hypernetwork, multiplying the incidence matrix by its transpose yields a cycle number matrix. Then, dividing each row of the cycle number matrix by the cycle number yields a cycle ratio matrix, and then adding each column to it gives the cycle ratio of each node. Thus, the new notion sheds some lights to future research on hypernetworks.

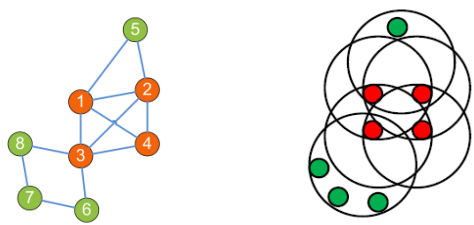

Figure 5. Left: an ordinary network; Right: its converted hypernetwork.

- **Brain functional networks**—Important roles of cliques and cavities

It was pointed out [9] that, although the human brain looks link sparsely connected, its clique structure therein is quite dense. It was found [9] that cliques play a very important role in cortical, visual and perceptive functions. However, from the conventional graph-theoretic viewpoint, one can only observe the connectivity among nodes, but not to discover deeper and higher-order structural characteristics of the brain, which needs more powerful mathematical tools such as algebraic topology. From an empirical research investigation, it was found [9] that cycles with longer girths are extremely important in the task of controlling the brain, and that the cavity structure is even more critical in the spreading patterns of the brain. Surprisingly, it was found [9] that the universal cavity structure in the brain does not exist in the conventional benchmark null network model, indicating the need of more powerful topological graph theory [14] beyond the classical algebraic graph theory in the studies of the brain.

- **Computational topology coming to network science**—Looking for higher-order topological features

Persistent homology[15], an important subject in algebraic topology, can be used to improve the computational accuracy in different spaces and to detect subtle details in a multi-scale space, recovering more essential features of a research object on the ground space. In contrast, the conventional techniques such as signal sampling and noise analysis as well as parameter selection may yield some false results. In a study of functional network formed by time-series data obtained by a weighted rank filtration technique, it was found [8] that cliques and cavities in a functional network have higher-order characteristics than the connectivity among nodes, which provides much more useful information, consistent with some existing studies [9]. By investigating the synchronization of Kuramoto oscillators using fMIR data, it was found [8] that persistent homology can reveal clearly some synchronous behaviors in the learning process of the brain, which were not discovered by conventional signal sampling and noise analysis. Typically, persistent homology help distinct strong and weak synchronization phenomena in communities of the brain network, and help detect functional changes through the learning process of the brain. A recent report [16] shows that persistent homology can be used to assist in topological data analysis, to reveal local, mesoscale and global properties and features of the network, using weighted, noisy and non-uniformly-sampled complex data, verified by EEG data analysis.

**Conclusion**

Using a sequence of clique vector spaces along with boundary operators to describe complex networks has well demonstrated that cliques, simplexes and fully-connected sub-networks are the backbones of various networks. This framework allows higher-level mathematical concepts and methods such as characteristic number, homology group and Betti number to play more significant roles in network science studies. They provide useful tools for uncovering and analyzing higher-order topological features and global structures of a complex network. Four representative classes of totally homogeneous networks have been examined, especially some elegant properties of the smallest $k$-cavity subnetworks and sync-optimal networks, revealing that cycle homogeneity is as important as node heterogeneity for understanding complex networks. Network synchronization criteria are originated from physics, which are then evolved via optimization to establishing the notion of totally homogenous networks, and finally connected to some invariants in algebraic topology, with significance demonstrated by brain research. This process highlights the interactions among network science, physics, biology and mathematics. When looking at an object from different angles, one finds different aspects about it. The situation is like what the famous Chinese poet Su Dongpo said, in his well-known poem [17], "From the side, a whole range; from the end, a single peak: Far, near, high, low, no two parts alike. Why can't I tell the true shape of Lu-shan? Because I myself am in the mountain." The new perspective of this paper hopefully would open up a new research direction in network science studies in the near future.

University Press, New York (1965).
18 Shi, D. H., Lü, L. & Chen, G. R. Supplementary Information. http://www.ee.cityu.edu.hk/~gchen/pdf/SLC-SI.pdf

**FUNDING**: This work is supported by the National Natural Science Foundation of China (Nos. 61174160, 11622538, 61673150), the Zhejiang Provincial Natural Science Foundation of China (No. LR16A050001), and the Hong Kong Research Grant Council under GRF Grant CityU11200317.